\documentclass{Interspeech}

\interspeechcameraready

\title{AC/DC: LLM-based Audio Comprehension via Dialogue Continuation}

\author[affiliation={1}]{Yusuke}{Fujita}
\author[affiliation={1}]{Tomoya}{Mizumoto}
\author[affiliation={1}]{Atsushi}{Kojima}
\author[affiliation={1}]{Lianbo}{Liu}
\author[affiliation={1}]{Yui}{Sudo}

\affiliation{}{SB Intuitions}{Japan}
\email{yusuke.fujita@sbintuitions.co.jp}

\usepackage{comment}

\usepackage{amsmath}
\usepackage{bm}
\usepackage[linesnumbered,ruled,vlined,commentsnumbered]{algorithm2e}
\usepackage{booktabs}
\usepackage{multirow}
\usepackage[table]{xcolor}
\usepackage{cite}
\usepackage{subfigure}

\def\*#1{\bm{#1}}

\begin{document}

\maketitle

\begin{abstract}
    
We propose an instruction-following audio comprehension model that leverages the dialogue continuation ability of large language models (LLMs). Instead of directly generating target captions in training data, the proposed method trains a model to produce responses as if the input caption triggered a dialogue. This dialogue continuation training mitigates the caption variation problem. Learning to continue a dialogue effectively captures the caption's meaning beyond its surface-level words. As a result, our model enables zero-shot instruction-following capability without multitask instruction tuning, even trained solely on audio captioning datasets. Experiments on AudioCaps, WavCaps, and Clotho datasets with AudioBench audio-scene question-answering tests demonstrate our model's ability to follow various unseen instructions.
\end{abstract}

\section{Introduction}

Automated audio captioning (AAC) is the task of describing audio content in natural language~\cite{Drossos2017_AAC, Mei2022_AAC}.
It has various potential applications, such as support systems for deaf and hard-of-hearing people to understand environmental sounds and video content retrieval systems to find desired sound scenes.
In this research field, acoustic scene classification (ASC) and audio event detection (AED) have been foundational tasks~\cite{Stowell2015_dcase}.
ASC addresses the problem of classifying the acoustic environment in an audio recording with predefined semantic labels, e.g., ``office'' and ``park''.
AED targets the identification of specific sound events, e.g., ``dog barking'', and detects when they occur.
Though ASC and AED provide categorical or timestamped event descriptions, their outputs are relatively coarse rather than a holistic narrative of the audio scene.
Therefore, AAC extends the tasks by offering free-form natural language summaries of audio content.
The DCASE Challenges have organized the AAC tasks and fostered advanced methods~\cite{Koizumi2020_dcase, Mei2021_dcase, Mei2022_dcase} with a standardized Clotho~\cite{Drossos2020_clotho} dataset.
The AAC task has been further extended to audio question-answering (AQA) that responds to various questions related to the audio content~\cite{Lipping2022ClothoAQAAC}.

One of the major problems in AAC and AQA is to handle variations in textual descriptions of the same audio content.
Whereas ASC and AED systems are designed to classify predefined labels with a small vocabulary, the free-form nature of audio captions requires large-vocabulary word sequences as training targets.
Compared to automatic speech recognition (ASR), which is a similar large-vocabulary task, AAC and AQA are more difficult for text-audio alignment.
While ASR transcriptions have fewer variations in their written form and are monotonically aligned to the corresponding audio, there are various audio caption candidates for describing the same audio content, such as ``{\it telephone bell is ringing''} versus ``{\it phone sounds are heard.''}
Moreover, the audio and the caption text are not always monotonically aligned, such as ``{\it a dog barking and then birds chirping,}'' versus ``{\it birds chirping after a dog barking.}''
Due to the lack of text uniqueness, the model training faces a data scarcity problem.

To address the text variation problem, prior work has explored the use of large-scale text generation models to augment training data.
WavCaps~\cite{Mei2024_wavcaps} introduced the large-scale audio captioning dataset, consisting of over 400K audio clips. In WavCaps, a large language model (LLM) transforms and filters the raw sound metadata to generate diverse and high-quality captions.
SLAM-AAC~\cite{chen2024slamaac} proposed a paraphrasing augmentation method based on back-translation with a machine translation model.
These approaches can be viewed as knowledge distillation from the text generation model into the AAC models.
Scaling up the training data using text generation models improves accuracy but requires a more significant training cost.

The knowledge of text generation models can be directly obtained by integrating LLMs into AAC and AQA models~\cite{deshmukh2023pengi,tang2024salmonn, ghosh2024gama, Tang2024_icassp, chen2024slamaac, Liu2024_interspeech}.
Such LLM-based models also benefit from instruction-following capability, which is critically important for the AQA task, answering various types of questions.
SALMONN~\cite{tang2024salmonn} integrates BEATs~\cite{Chen2023_beats} and Whisper~\cite{Radford2023_whisper} encoders with Vicuna LLM~\cite{Chiang2023_vicuna} and is trained across multiple tasks (including AAC, AQA, and many speech and music tasks) with task-specific prompts.
Researchers examined various audio encoders such as CED~\cite{Dinkel2024_icassp} and EAT~\cite{Chen2024_eat} for improving the AAC performance~\cite{Liu2024_interspeech, chen2024slamaac}.
Though multitask-trained models like SALMONN can follow various instructions, they struggle with making variations in tasks and vocabulary used in training data.
On the other hand, AAC-specific models like SLAM-AAC~\cite{chen2024slamaac} give a significant boost in AAC performance, but they often lack the instruction-following capability.
In all the aforementioned models, sufficient variation of audio-text pairs for each task in training data is crucial.
In other words, transferred knowledge from LLMs is limited to the distribution of target text in training data. 
As a result, the existing models cannot follow {\it unseen task} instructions.

\begin{figure*}[t]
  \begin{center}
  \includegraphics[scale=0.8]{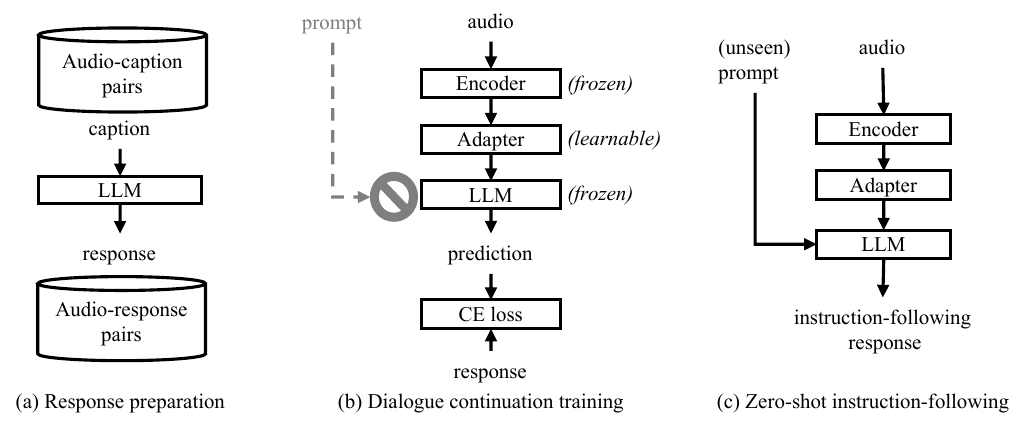}
  \end{center}
  \caption{System diagram of proposed dialogue continuation training.}
  \label{fig:systems}
\end{figure*}

In this paper, we propose a training method for audio comprehension models that follow unseen instructions with no task-specific training data.
Unlike conventional AAC models that learn to produce captions, the proposed method trains a model to produce dialogue responses.
The response texts for the model training targets are prepared by feeding the ground-truth captions to the backbone LLM without any task-specific instructions.
The trained model produces a response as if the audio caption triggered a dialogue.
This dialogue continuation training prevents overfitting to the caption vocabulary and the captioning task itself.
%The {\it indirect} learning scheme captures the caption's meaning rather than the caption's surface-level words.
Since the non-instructed training targets are generated by the backbone LLM itself, the trained models are not biased to any task-specific prompts and are ready to fully leverage the instruction-following capability of the LLM.
As a result, the trained model answers various audio-related questions without either task-specific data augmentation or multitask training data.
Experiments on AudioCaps~\cite{kim2019-audiocaps}, Clotho~\cite{Drossos2020_clotho}, WavCaps~\cite{Mei2024_wavcaps} datasets with AQA testsets in AudioBench~\cite{wang2024_audiobench} demonstrate the zero-shot instruction-following ability.

\section{Proposed Method}

Our method trains a model without explicit task prompts. Instead, we use LLM-generated responses as targets. Due to this dialogue continuation training, we can append unseen task prompts to the frozen LLM in the inference phase, yielding instruction-following responses.
Figure~\ref{fig:systems} shows the overview of the proposed method.

\subsection{Baseline audio captioning model}
\label{sec:baseline}

Our model architecture is based on SLAM-AAC~\cite{chen2024slamaac}, which is an LLM-based AAC model. The model uses a pretrained audio encoder, an adapter based on linear layers, and an instruction-tuned LLM decoder.
We slightly modify the SLAM-AAC architecture so as to consider both prefix and suffix prompts as inputs to the LLM decoder, whereas SLAM-AAC only considers the suffix prompt.

In our {\it baseline} model, a $T$-length input audio sequence $\*x \in \mathbb{R}^T$ is transformed into a $L$-length caption sequence $\*y \in \mathcal{V}^L$, conditioned on prefix $\*p$ and suffix $\*s$ prompt texts:
\begin{align}
    \*X &= \mathsf{Encoder}(\*x) \in \mathbb{R}^{D' \times T'}, \\
    \*A &= \mathsf{Adapter}(\*X; \Theta) \in \mathbb{R}^{D \times T'}, \\
    \*P &= \mathsf{Tokenizer}(\*p) \in \mathbb{R}^{D \times P}, \\
    \*S &= \mathsf{Tokenizer}(\*s) \in \mathbb{R}^{D \times S}, \\
    \*C &= [\*P; \*A; \*S] \in  \mathbb{R}^{D \times (P+T'+S)}, \\
    p(\*y_i \mid \*y_{<i},\*C) &= \mathsf{LLM}(\*y_{<i},\*C) \in [0,1]^{|\mathcal{V}|} \label{eq:llm} \\
    & \qquad\qquad\qquad (1 \le i \le L), \nonumber
\end{align}
where $\mathcal{V}$ is a text token vocabulary, $\mathsf{Encoder}$ is the audio encoder function to audio embeddings $\*X$, $D'$ is the dimension of an audio embedding, $T'$ is the length of audio embeddings, $\Theta$ is learnable parameters of the linear layers in the $\mathsf{Adapter}$ function transforming $\*X$ into $\*A$, $D$ is the embedding dimension matching to the LLM token embedding, the $\mathsf{Tokenizer}$ function tokenizes a prompt text sequence ($\*p$ or $\*s$) and generates corresponding token embeddings, $P$ and $S$ are the lengths of tokenized embeddings,
and $\mathsf{LLM}$ is the LLM decoder that autoregressively produces a distribution of the $i$-th output token $\*y_i \in \mathcal{V}$ conditioned on the prompt token embeddings $\*C$ and previous output tokens $\*y_{<t}$.
We use beam search to obtain the best sequence $\*y$ from Eq.~\ref{eq:llm}, and the output sequence length $L$ is determined when the end-of-sentence token is generated.

To prepare prefix and suffix prompts, we use the chat template\footnote{https://huggingface.co/docs/transformers/chat\_templating} with ``system'', ``user'', and ``assistant'' roles. We use the fixed prompt specific to the AAC task: ``{\it Describe the audio you hear},'' in the system prompt for the baseline model.
The audio embeddings $\*A$ are placed in the user prompt, and then the assistant prompt is appended for generating the output caption from the assistant role.
The learnable parameters $\Theta$ are only in the adapter.
They are learned to minimize cross-entropy loss between the predicted distribution and the ground-truth caption in the AAC training data $\*y^*_i$:
\begin{align}
    \mathcal{L}_\mathsf{caption} = - \frac{1}{|\*y^*|} \sum_{i=1}^{|\*y^*|} \log p(\*y^*_i \mid \*y^*_{<i}, \*C).
\end{align}

\subsection{Dialogue continuation training}
\label{sec:dct}
Instead of using ground-truth audio captions, the proposed model replaces the training targets with responses to the audio caption.
As depicted in Figure~\ref{fig:systems}(a), we prepare the response to the caption in training data.
The response $\*r^*$ is generated by feeding the ground-truth audio caption $\*y$ to the same $\mathsf{LLM}$ function used in the baseline model:
\begin{align}
  \*Y^* &= \mathsf{Tokenizer}(\*y^*), \\
  \*P' &= \mathsf{Tokenizer}(\*p^\mathsf{(noinst)}), \\
  \*S' &= \mathsf{Tokenizer}(\*s^\mathsf{(noinst)}), \\
  \*C' &= [\*P'; \*Y^*; \*S'], \\
  p(\*r_i^* | \*r_{<i}^*, \*C') &= \mathsf{LLM}(\*r_{<i}^*, \*C').
\end{align}
Here, $\*p^\mathsf{(noinst)}$ and $\*s^\mathsf{(noinst)}$ are the prompt texts with no specific instruction but only contain special tokens indicating start and end markers of user and assistant messages.
The best sequence $\*r^*$ is obtained through temperature-based sampling to encourage variations of the training targets.

The response $\*r^*$ is different from the one obtained from conventional paraphrasing augmentation~\cite{chen2024slamaac}.
Whereas paraphrasing augmentation inevitably increases the amount of training data, our method does not change the data amount but rather increases task variations.
For example, a paraphrasing of ``{\it In the forest a flock of birds is chirping.}'' is ``{\it In the forest, a \textbf{group} of birds are chirping.}''
In contrast, the proposed response is ``{\it What a lovely scene! The forest is alive with the sweet sounds of birds chirping away!}''
With no specific instruction, we obtain a wide variety of dialogue-like responses capturing the meaning of audio content.
Note that we use only a single sampled response per a ground-truth caption, which is sufficient to obtain instruction-following capability in our experiments.

The proposed training process is shown in Figure~\ref{fig:systems}(b).
Given the prepared response $\*r^*$, the adapter's parameters $\Theta$ are optimized to minimize cross-entropy loss for the prepared responses instead of captions:
\begin{align}
    \mathcal{L}_\mathsf{DCT} = - \frac{1}{|\*r^*|} \sum_{i=1}^{|\*r^*|} \log p(\*r^*_i \mid \*r^*_{<i}, \*C').
\end{align}
Since the prepared target is obtained through the same LLM as used in the model, the LLM parameters can be frozen, and the adapter training can focus on extracting the caption's meaning rather than surface-level words to produce the same response as if the caption is fed into the LLM.

\subsection{Zero-shot instruction-following inference}

Being trained with no task-specific instruction as described in Sec.~\ref{sec:dct}, the adapter does not overfit to the audio captioning task.
Therefore, the proposed model can accept various instructions in the inference phase.
By setting $\*p^\mathsf{(inst)}$ and $\*s^\mathsf{(inst)}$ appropriately according to the chat template for the backbone LLM, the response $\*r$ follows the specified instruction:
\begin{align}
  \*P'' &= \mathsf{Tokenizer}(\*p^\mathsf{(inst)}), \\
  \*S'' &= \mathsf{Tokenizer}(\*s^\mathsf{(inst)}), \\
  \*C'' &= [\*P''; \*A; \*S''], \\
  p(\*r_i | \*r_{<i}, \*C'') &= \mathsf{LLM}(\*r_{<i}, \*C'').
\end{align}
For example, if we use the prompt, ``Describe the audio content.'' the response $\*r$ does the AAC task.
Using prompts like ``Does the sound have speech?" the model answers the audio-related question.

%Note that models trained with various multiple tasks can follow various instructions in a similar manner. However, such models need large amount of training data and some dataset balancing among the training tasks. Unlike such multitask models, the proposed model is solely trained on the single-task audio captioning dataset with no task-specific instruction, resulting in keeping the instruction-following capability that the backbone LLM has already acquired.

\subsection{Relation to prior work}

The ASR-oriented dialogue continuation has been investigated in AudioChatLlama~\cite{fathullah-etal-2024-audiochatllama}.
Similar to the proposed method, it targets LLM responses without task prompts using ASR training data, i.e., transcriptions.
The audio-text alignment in ASR is relatively easier than the AAC task.
Moreover, the ASR-oriented dialogue continuation model captures the speech pronunciation rather than the meaning, leading to difficulty in recognizing rare words, as mentioned in \cite{fathullah-etal-2024-audiochatllama}.
On the other hand, our AAC-oriented model deals with the situation in which the same audio feature could be assigned to different words but with the same meaning.
Our work is the first attempt to obtain audio-text alignment for non-speech audio data using dialogue continuation.

\section{Experiments}

\subsection{Experimental Setup}
\subsubsection{Data}

The main training dataset in the experiments was AudioCaps~\cite{kim2019-audiocaps}.
We also used WavCaps~\cite{Mei2024_wavcaps} and Clotho~\cite{Drossos2020_clotho} as additional training data for validating the scalability of the proposed method.
We did not apply paraphrasing augmentation in SLAM-AAC~\cite{chen2024slamaac} to check if the proposed method can mitigate the data scarcity problem without expanding data volume.

The evaluation metrics and datasets were derived from AudioBench~\cite{wang2024_audiobench}.
We used their AQA benchmarks, including Clotho-AQA, AudioCaps-QA, and WavCaps-QA, to validate the models' ability to follow various instructions. The AQA accuracy was judged by Llama-3-70B-Instruct according to the AudioBench framework.
We also used test sets of AudioCaps to evaluate the AAC performance, which is the originally intended task of the baseline model and the training dataset.

\subsubsection{Modeling}

We developed our baseline model using SLAM-AAC~\cite{chen2024slamaac} with slight modifications described in Sec.~\ref{sec:baseline}.
The baseline model used EAT-base\footnote{The slam-aac recipe in https://github.com/X-LANCE/SLAM-LLM} as the audio encoder pretrained and finetuned on AudioSet~\cite{Gemmeke2017_audioset} dataset.
The adapter was a 2-layer linear projection with 5x downsampling.
We used Llama-3-8B-Instruct\footnote{https://huggingface.co/meta-llama/Meta-Llama-3-8B-Instruct} as the backbone LLM.
We did not use the CLAP-Refine strategy in SLAM-AAC~\cite{chen2024slamaac} for simplicity.
The training targets of the proposed model were prepared as described in Sec.\ref{sec:dct} using the backbone LLM with a temperature of 0.7. 

All the models were optimized with a batch size of 4. We ran a 1,000-iteration warmup to a peak learning rate of 1e-4 before linearly decaying over 100,000 iterations.
We conducted the model validation steps every 1,000 iterations,
Though the backbone LLM was frozen, we examined with and without applying LoRA~\cite{hu2022lora} finetuning in the LLM decoder.

\subsubsection{Inference}

In the inference phase, we used beam search with a beam size of 4.
The instruction prompt for the model varies between tasks.
For the AQA tasks, we used this instruction:

\fbox{%
    \parbox{0.9\linewidth}{%
"system": "Answer the question provided after the audio caption within 10 words." \\
"user": "\{AUDIO\}. Question: \{INSTRUCTION\},"

    }%
}
where "\{AUDIO\}" is the placeholder replaced with the audio embedding, and "\{INSTRUCTION\}" is replaced with the actual question in the test dataset.
For the AAC task, we used the original prompt used for the baseline model training:

\fbox{%
    \parbox{0.9\linewidth}{%
"system": "Describe the audio content in one 10-word sentence." \\
"user": "\{AUDIO\}."
    }%
}

\subsection{Results}

\begin{table*}[t]
    \centering
    \caption{Performance metrics for different training configurations. Each accuracy was judged by Llama-3-70B-Instruct and rescaled to a 100-point scale following the method in AudioBench~\cite{wang2024_audiobench}.
    Models C1 and C2 were from AudioBench paper~\cite{wang2024_audiobench}.
    }
    \vspace{-2.5mm}
    \begin{tabular}{llcc|rrrr|rr}
        \toprule
        & \multicolumn{3}{c|}{\bf Training configuration} & \multicolumn{4}{c|}{\textbf{AQA accuracy}} & \multicolumn{1}{l}{\textbf{AAC accuracy}}\\
        \textbf{ID} & \textbf{Data} & \textbf{Target} & \textbf{LoRA} & \textbf{Clotho} & \textbf{AudioCaps} & \textbf{WavCaps} & \textbf{Avg.} & \textbf{AudioCaps} \\
        \midrule
        A1 & \multirow{4}{*}{AudioCaps} & caption & N & 28.56 & 39.74 & 29.61 & 32.64 & \bf 51.43  \\
        A2 & & proposed & N & \bf 52.88 & \bf 47.54 & \bf 42.37 & \bf 47.60 & 46.20 \\ \cmidrule{3-9}
        A3 & & caption & Y & 32.38 & 41.85 & 35.20 & 36.48 & \bf 52.55   \\
        A4 & & proposed & Y & \bf 56.38 & \bf 47.09 & \bf 37.96 & \bf 47.14 & 43.13 \\
        \midrule
        B1 & \multirow{4}{6em}{AudioCaps +WavCaps +Clotho} & caption & N & 29.47 & 39.42 & 30.26 & 33.05 & \bf 51.27 \\
        B2 & & proposed & N & \bf 55.82 & \bf 47.99 & \bf 39.28 & \bf 47.70 & 45.53 \\ \cmidrule{3-9}
        B3 & & caption & Y & 26.73 & 42.56 & 31.58 & 33.62 & \bf 52.08\\
        B4 & & proposed & Y & \bf 54.68 & \bf 47.73 & \bf 40.59 & \bf 47.67 & 43.51 \\
        \midrule
        C1 &\multicolumn{3}{c|}{SALMONN \cite{tang2024salmonn}} & 51.18 & 47.03 & 46.25 & 48.15 & 34.37 \\
        C2 &\multicolumn{3}{c|}{Qwen-Audio-Chat~\cite{chu2023_qwenaudio}} & 58.20 & 47.99 & 38.68 & 48.29 & 47.99 \\
        \bottomrule
    \end{tabular}
    \label{tab:results}
\end{table*}

\begin{table*}[t]
    \centering
    \caption{Examples of instruction-following test. The input audio is ``river\_mouth3.wav'' in Clotho-AQA. The ground-truth caption is ``body of water splashes around over and over.'' The ground-truth answer to the question ``Are there waves?'' is ``Yes, waves do exist.''
    }
    \vspace{-2.5mm}
    \begin{tabular}{l p{65mm} p{70mm}}
        \toprule
        \textbf{Model} & \textbf{Instruction} & \textbf{Prediction} \\
        \midrule
        baseline (A3) & (noinst) & Water splashing and gurgling as wind blows into a microphone \\ \cmidrule{2-3}
        & Describe the audio content in one 10-word sentence. & Water splashing and gurgling \\ \cmidrule{2-3}
                 & Are there waves? & Water splashing and gurgling are present, the wind is blowing, and birds are chirping in the background \\ \cmidrule{2-3}
                 & Describe the audio content in one 10-word {\bf German} sentence. & Water splashing and gurgling as wind blows into a microphone \\ \midrule
        proposed (A4) & (noinst) & The soothing sound of waves crashing against the shore! It's a calming and peaceful scene, isn't it?  \\ \cmidrule{2-3}
        & Describe the audio content in one 10-word sentence. & Waves crashing against the shore create soothing ocean soundscape ambiance. \\ \cmidrule{2-3}
                 & Are there waves? & Yes, there are waves! \\ \cmidrule{2-3}
                 & Describe the audio content in one 10-word {\bf German} sentence. & Das Wasser plätschert und schwappt rhythmisch am Strand entlang. \\
        \bottomrule
    \end{tabular}
    %\vspace{-1mm}
    \label{tab:example}
\end{table*}

Table~\ref{tab:results} shows the performance metrics for different models.
We observed that the proposed training target achieved much better AQA accuracy than the conventional training target (caption) for all AQA benchmarks by large margins (A1 vs. A2, A3 vs. A4).
The LoRA finetuning demonstrated improvement when trained with the conventional target (A1 vs. A3), while no consistent behaviors were observed when trained with the proposed target.
Our best result on the averaged AQA accuracy (47.70\%; B2) was comparable with that of Qwen-Audio-Chat (48.29\%; C2). The results indicate that the proposed method is efficient in acquiring AQA capability without large-scale multitask data augmentation.

For the AAC task, the baseline models were better than the proposed models. It was expected since the proposed model was built for general audio comprehension and does not limit their vocabulary to that of baseline AAC models.
Our AAC results were slightly worse than that of Qwen-Audio-Chat trained with a large-scale multitask setup including AAC targets in training data.
The results suggest that the proposed model is more general and task-agnostic than other models and could be further finetuned for specified purposes using task-specific targets. We leave this direction for future work.

We observed similar trends when trained on more data (B1-B4), but there was no significant improvement (vs. A1-A4). This could be due to the relatively few learnable parameters in our models. Finetuning the audio encoder jointly with the adapter may improve performance.

Table~\ref{tab:example} demonstrates examples of instruction-following results.
The baseline model (A3) did not follow instructions and produced similar captions.
In contrast, the proposed model (A4) followed various instructions in a zero-shot manner. With no instruction, it produced a dialogue-like response to the audio.
The results suggest that the proposed method effectively utilizes the original instruction-following capability in the backbone LLM.

\section{Conclusion}
We proposed an audio comprehension model that enhances audio QA by leveraging LLM's dialogue continuation capability.
Experiments demonstrated its effectiveness and potential for flexible audio understanding.
In future work, we plan to combine dialogue continuation training with explicit instruction-tuning to further enhance task-specific performance while preserving zero-shot capability.
\bibliographystyle{IEEEtran}
\bibliography{mybib}

\end{document}